\documentclass{ws-procs975x65}
\def\beq{\begin{equation}}
\def\eeq{\end{equation}}
\def\beqn{\begin{eqnarray}}
\def\eeqn{\end{eqnarray}}
\newcommand{\be}{\begin{eqnarray}}
\newcommand{\ee}{\end{eqnarray}}

\begin{document}

\title{SUPER-RENORMALIZABLE GRAVITY}

\author{Leonardo Modesto}
\address{Department of Physics, Fudan University \& Center for Field Theory and Particle Physics,\\
Shanghai,  200433, China\\
E-mail: lmodesto@fudan.edu.cn} 

\begin{abstract}

We review a class of higher derivative theories of gravity consistent at quantum level.
This class is marked by a non-polynomal entire function (form factor), which averts
extra degrees of freedom (including ghosts) and improves the high energy
behaviour of the loop amplitudes. By power counting arguments, it is
proved that the theory is super-renormalizable, i.e. only one-loop divergences survive. 
At classical level, black holes and cosmological solutions are singularity free. 

\end{abstract}


\bodymatter
\vspace{0.6cm}

\hspace{-0.3cm} We hereby show that is possible to find a solution of the quantum gravity puzzle in the
quantum field theory framework. We know that all the other fundamental interactions are consistent 
with quantum mechanics and perturbation theory. 
The higher derivative theory of gravity \cite{Stelle, shapiro} quoted here 
\cite{modesto, BM, M2, M3, Krasnikov, Tombo}
is compatible with the two properties above at the same level of the other three fundamental interactions. 
The theory is renormalizable and unitary, but it is non-polynomial due to the operators 
$R \, \sum_{r=0}^{+ \infty} c_r \Box^r \, R$ and 
 $R_{\mu \nu} \, \sum_{r=0}^{+ \infty} c_r \Box^r \, R^{\mu \nu}$
that make the Lagrangian non-local ($\Box$ is the covariant D'Alembertian operator). 
As it is well known, non-locality is a property shared by all the other fundamental interactions when the one-loop effective action is considered. 
%

Let us now explore the mathematical details of the theory.
In a $D$-dimensional spacetime the most general action reads 
\be
\hspace{-0.2cm} 
 \mathcal{L}_D \! =  \! 
 \frac{2 a_{\kappa}}{\kappa^{2} }  R+ R_{\mu \nu}  \gamma_2(\Box) R^{\mu \nu} \! + R \,\gamma_0( \Box)  R 
+ R_{\mu \nu \rho \sigma} \gamma_4(\Box) R^{\mu \nu \rho \sigma} 
+ 
R^3 + \dots + R^{\frac{D}{2}} 
 ,  \label{Daction}
\ee
where 
the three ``form factors" $\gamma_{0,2,4}(\Box)$ are ``non-polynomial trascendental entire functions" of the covariant D'Alembertian operator. 
In $D=4$ the Lagrangian (\ref{Daction}) simplifies to just the first three operators in (\ref{Daction})
with the following clear definitions 
\be 
\hspace{0.1cm} 
\mathcal{L} = 
  \frac{2 a_{\kappa}}{\kappa^{2} } R+
  b_0 R_{\mu \nu}   
  R^{\mu \nu} +
  a_0 R^2 + 
   R_{\mu \nu}   \, h_2(-\Box_{\Lambda})
  R^{\mu \nu}  + R \,   h_0(-\Box_{\Lambda}) 
  R \, . 
   \label{4Daction}
\ee 
In (\ref{4Daction}) the trascendental entire functions $h_i(-\Box_{\Lambda})$ will be defined later; 
$\Box_{\Lambda} := \Box/\Lambda^2$, $\Lambda$ is an invariant mass scale; $a_{\kappa}$, $a_0$, $b_0$ 
are coupling constants subjected to quantum renormalizations; 
 $\kappa^2 = 32 \pi G_N$. 
To address the unitarity problem we first calculate the graviton propagator. 
%
The gauge invariant part largely simplifies to the following form, 
%
%
\be
 && \hspace{-0.75cm} 
 \mathcal{O}^{-1}(k)
 = \frac{1}{k^2}
\left( \frac{P^{(2)}}{\bar{h}_2} 
- \frac{P^{(0)}}{2\bar{h}_0 } \right) 
\, ,  \,\,\,\, 
\bar{h}_2 := a_{\kappa} + \frac{\kappa^2 \Lambda^2}{2} z b_0 + \frac{\kappa^2 \Lambda^2}{2} z h_2(z) \,,  \nonumber \\
&& \hspace{-0.75cm} 
\bar{h}_0 := a_{\kappa}  - \kappa^2 \Lambda^2 z (b_0+h_2(z)) - 3 \kappa^2 \Lambda^2 z (a_0+ h_0(z)) .
\label{propgauge}
\ee
%
%
It is easy to see that 
any polynomial choice, $p_n(z)$, for the
the functions $h_2$ and $h_0$ will introduce $n$-additional poles from the $n$ zeros in $\bar{h}_i (i=0,2)$.
The residues of some of these poles will be negative, consequently introducing ghosts in the spectrum and  
then causing loss of unitarity.
To overcome this issue, we demand 
the following general properties for the transcendental entire functions $h_i(z)$ ($i = 0,2$) and/or 
$\bar{h}_i(z)$ ($i = 0,2$)
\cite{Tombo}:
\begin{enumerate}
\renewcommand{\theenumi}{(\roman{enumi})}
\renewcommand{\labelenumi}{\theenumi}
\item $\bar{h}_i(z)$ ($i=0,2$) is real and positive on the real axis and it has no zeroes on the 
whole complex plane $|z| < + \infty$. This requirement implies that there are no 
gauge-invariant poles other than the transverse massless physical graviton pole.
\item $|h_i(z)|$ has the same asymptotic behavior along the real axis at $\pm \infty$.
\item There exists $\Theta>0$ such that 
\be
&& \lim_{|z|\rightarrow + \infty} |h_i(z)| \rightarrow | z |^{\gamma} \, , \,\,\,  \gamma \geqslant 2 \,, 
\label{tombocond}
\ee 
for the argument of $z$ in the following conical regions  
\be
&& \hspace{-0.2cm} 
C = \{ z \, | \,\, - \Theta < {\rm arg} z < + \Theta \, , \,\,  \pi - \Theta < {\rm arg} z < \pi + \Theta \} , \hspace{0.2cm} 
{\rm for } \,\,\, 0< \Theta < \pi/2. \nonumber 
\ee
This condition is necessary in order to achieve the supe-renormalizability of the theory that we 
are going to show here below. The necessary 
asymptotic behavior is imposed not only on the real axis, (ii) but also in the conic regions that surround it.  
In an Euclidean spacetime, the condition (ii) is not strictly necessary if (iii) applies.
\end{enumerate}

Imposing the conditions (i)-(iii) we have the option of choosing the following form for the functions 
$h_i$,
\be
h_2(z) = 2 \tilde{a}_{\kappa} \frac{ V(z)^{-1} -1}{\kappa^2 \Lambda^2 z} - \tilde{b}_0 \,\, , \hspace{0.2cm} 
h_0(z) = - \tilde{a}_{\kappa} \frac{V(z)^{-1} -1}{
\kappa^2 \Lambda^2 \, z} + \tilde{a}_0\, , 
\label{hzD}
\ee
for general parameters $\tilde{a}_{\kappa}$, $\tilde{a}_0$ and $\tilde{b}_0$. 
At quantum level the coupling constants $a_{\kappa}(\mu)$, 
$a_0(\mu)$ and $b_0(\mu)$ are renormalized at some scale $\mu$.
If we assume the theory to be renormalized at a certain scale $\mu_0$ and we identify 
 $\tilde{a}_{\kappa} = a_{\kappa}(\mu_0)$, 
$\tilde{a}_0 = a_0(\mu_0)$, $\tilde{b}_0 =b_0(\mu_0)$, then the propagator simplifies to 
\be
 \hspace{-0.75cm} \mathcal{O}^{-1}(k)
 = \frac{V(k)}{2 k^2}
\left( 2 P^{(2)} - P^{(0)}  \right) ,
\ee
and no other poles are manifest besides the graviton one. If we choose another renormalization scale, the bare propagator acquires poles that will cancel out with a self-energy shift in the dressed propagator.

Here we offer an explicit  example of how a trascendental entire function
$V(z)^{-1}$ does not introduce extra poles in the graviton propagator and satisfies the properties (i)-(iii).  
If we define $V(z)^{-1} := \exp H(z)$, then a possible choice for 
the entire function $H(z)$ is 
\be
 H(z) = \frac{1}{2} \left[ \gamma_E + 
\Gamma \left(0, p_{\gamma+1}^{2}(z) \right)  + \log \left( p^2_{\gamma+1}(z) \right) \right]  
\, , 
\,\,\,\,\,
{\rm Re}( p_{\gamma+1}^{2}(z) ) > 0,
\label{HD}
\ee
where $p_{\gamma+1}(z)$ is a polynomial of degree $\gamma+1$ (for example $p_{\gamma+1} = z^{\gamma+1}$.)

Let us eventually examine under what conditions the theory result to be power counting renormalizable. 
According to the property (iii) in the high energy regime, the propagator in the momentum space goes as 
$\mathcal{O}^{-1}(k) \sim 1/k^{2 \gamma +4}.$ 
The $n$-graviton interaction has the 
same leading scaling of the kinetic term, 
and the superficial degree of divergence in a four dimensional spacetime reads 
\be
\delta_{\rm G} \leqslant 4 L - (2 \gamma +  4) I + (2 \gamma + 4) V 
 = 4 - 2 \gamma  (L - 1). \nonumber 
\ee
It follows that the higher divergence we can find in the loop expansion is $\delta_{\rm G} =4$ for $L=1$.
For $L>1$ and $\gamma >2$ the loop amplitudes are convergent. 

So far we considered the theory at quantum level. But what about the classical solutions?
After the simplifications explained throughout the paper, the four dimensional action reduces to 
\be
\hspace{0.55cm}
\boxed{
\begin{array}{rcl}
&& \hspace{-0.2cm}
\mathcal{L} =  2 \kappa^{-2} \sqrt{|g|} \Big( R  
- G_{\mu\nu} \,  \frac{ e^{H(-\Box_{\Lambda}) }-1}{\Box}
\,  
R^{\mu\nu} \Big) 
\end{array}
}
\ee
and the equations of motion at the order $O(R^2)$ simplify to 
\be
 G_{\mu \nu} + O(R^2) = 8 \pi G_N \,  e^{- H(-\Box_{\Lambda})} T_{\mu \nu} \, . 
\ee
We can solve the above equations in the case of a spherically symmetric spacetime, obtaining regular black hole solutions \cite{ModestoMoffatNico, modesto, M2}. The theory also admits 
bouncing cosmological exact solutions
\cite{M5, Bis2, koshe}.

\end{document}